\documentclass[conference,letterpaper]{IEEEtran}
\usepackage{enumerate}

\usepackage{mathtools}
\usepackage{float}
\usepackage[center,singlelinecheck=false]{caption}
\usepackage{cite}
\usepackage{graphicx} %package to manage images
\usepackage{epstopdf}
\usepackage{amsmath} % assumes amsmath package installed
\usepackage{amssymb}  % assumes amsmath package installed
\usepackage{amsthm}
\usepackage[linesnumbered,ruled,vlined]{algorithm2e} % ,noend %remove "vlined" for puting "end"
\usepackage{nicematrix}

\usepackage[utf8]{inputenc}
\usepackage{color}
\usepackage{multirow}
\usepackage{optidef}
\usepackage[bottom=0.980in, right=0.63in, left=0.63in, top=0.72in]{geometry}

\theoremstyle{definition}
\newtheorem{example}{Example}

\newtheorem{lemma}{Lemma}

\newtheorem{definition}{Definition}

\newtheorem{remark}{Remark}

\DeclareMathOperator{\bin}{bin}
\DeclareMathOperator{\supp}{supp}

\DeclareMathOperator{\w}{w}

\newcommand{\wm}{\mathrm{w}_{\min}}
\newcommand{\dm}{d_{\min}}
\newcommand{\Awm}{A_{\mathrm{w}_{\min}}}

\newcommand{\Aiwm}{A_{i,\mathrm{w}_{\min}}(\I)}

\newcommand{\Ki}{\mathcal{K}_i}

\newcommand{\B}{\mathcal{B}}

\newcommand{\I}{\mathcal{I}}

\renewcommand{\H}{\mathcal{H}}

\newcommand{\J}{\mathcal{J}}

\newcommand{\F}{\mathcal{F}}
\newcommand{\M}{\mathcal{M}}

\newcommand{\C}{\mathcal{C}}
\newcommand{\Ci}{\mathcal{C}_i}
\newcommand{\CI}{\mathcal{C}(\mathcal{I})}
\newcommand{\CiI}{\mathcal{C}_i(\mathcal{I})}

\newcommand{\D}{\mathcal{D}}

\newcommand{\R}{\mathcal{R}}

\newcommand{\EI}{\text{E}_{\text{I}}}
\newcommand{\EII}{\text{E}_{\text{II}}}

\newcommand{\bG}{\mathbf{G}}

\newcommand{\bI}{\mathbf{I}}

\newcommand{\be}{\mathbf{e}}

\newcommand{\br}{\mathbf{r}}

\newcommand{\bc}{\mathbf{c}}

\newcommand{\bzero}{\mathbf{0}}

\newcommand{\bgi}{\mathbf{g}_i}
\newcommand{\bgj}{\mathbf{g}_j}
\newcommand{\bgm}{\mathbf{g}_m}
\newcommand{\bgf}{\mathbf{g}_f}
\newcommand{\bgh}{\mathbf{g}_h}
\newcommand{\bg}{\mathbf{g}}
\newcommand{\bp}{\mathbf{p}}
\newcommand{\bGN}{\boldsymbol{G}_N}
\newcommand{\bd}{\mathbf{d}}
\newcommand{\bq}{\mathbf{q}}
\newcommand{\bu}{\mathbf{u}}

\newcommand{\ft}{\mathbb{F}_2}

\usepackage{xcolor}
\def\BibTeX{{\rm B\kern-.05em{\sc i\kern-.025em b}\kern-.08em
    T\kern-.1667em\lower.7ex\hbox{E}\kern-.125emX}}

\usepackage{fancyhdr}

\usepackage[compact]{titlesec}
\titlespacing{\section}{0pt}{*0.8}{*0.8}
\titlespacing{\subsection}{0pt}{*0.6}{*0.6}
 
\IEEEoverridecommandlockouts                              % This command is only
\title{
PAC Codes Meet CRC-Polar Codes 
\vspace{-3pt}
}

\author{
\IEEEauthorblockN{Xinyi Gu}
\IEEEauthorblockA{%Faculty of Exact Sciences,\\
University of New South Wales\\ 
Sydney, Australia\\ 
Email: xinyi.gu@student.unsw.edu.au \vspace{-15pt}}
\and
\IEEEauthorblockN{Mohammad Rowshan}
\IEEEauthorblockA{%School of Electrical Engineering and Telecommunications,\\
University of New South Wales\\ 
Sydney, Australia\\ 
Email: m.rowshan@unsw.edu.au \vspace{-15pt}}
\and

\IEEEauthorblockN{Jinhong Yuan}%, {\em Fellow, IEEE}}
\IEEEauthorblockA{%School of Electrical Engineering and Telecommunications,\\
University of New South Wales\\ 
Sydney, Australia\\ 
Email: j.yuan@unsw.edu.au \vspace{-15pt}}
\vspace{-15pt}
} 

\begin{document}
\maketitle
\pagestyle{plain}
\thispagestyle{fancy}
\fancyhf{}
% \lhead{\color{blue} 
% %\footnotesize
% To be presented at the 2024 IEEE International Symposium on Information Theory (ISIT 2024)\\
% Session: TH4.R9: Topics in Modern Coding Theory 3
% }
\cfoot{}

%%%%%%%%%%%%%%%%%%%%%%%%%%%%%%%%%%%%%%%%%%%%%%%%%%%%%%%%%%%%%%%%%%%%%%%%%%%%%%%%
\begin{abstract}
CRC-Polar codes under SC list decoding are well-regarded for their competitive error performance. This paper examines these codes by focusing on minimum weight codewords, breaking them down into the rows of the polar transform. Inspired by the significant impact of parity check bits and their positions, we apply a shifted rate-profile for polarization-adjusted convolutional (PS-PAC) codes, thereby achieving similar improvements in the weight distribution of polar codes through precoding. The results demonstrate a significant improvement in error performance, achieving up to a 0.5 dB power gain with short PS-PAC codes. 
Additionally, leveraging convolutional precoding in PAC codes, we adopt a continuous deployment (masking) of parity check bits derived from the remainder of continuous division of the partial message polynomial and the CRC polynomial over frozen positions in the rate-profile. This approach enhances performance for medium-length codes, with an overall improvement of 0.12 dB. %Notably, the proposed schemes do not employ any search or optimization algorithms. \mohammad{I wrote this before reading your manuscript based on raw view without using your vocabularies. Feel free to improve it and add more details, in particular about the results. Do not make it long, please. I will slowly read your manuscript today (not all of it).}
\end{abstract}

%%%%%%%%%%%%%%%%%%%%%%%%%%%%%%%%%%%%%%%%%%%%%%%%%%%%%%%%%%%%%%%%%%%%%%%%%%%%%%%%

\section{INTRODUCTION}
\label{sec:intro}

Polar codes \cite{arikan2009channel} are provably a class of capacity-achieving codes. However, they do not provide satisfactory error correction performance under a relatively low complexity successive cancellation (SC) decoding in finite block length. To address this drawback, SC list (SCL) decoding  \cite{tal2015list} provides a near maximum likelihood (ML) block error rate (BLER) at the cost of high computational complexity. Alternatively, error performance can be enhanced through pre-transformation. For instance, concatenating polar codes with cyclic redundancy check (CRC) codes (CRC-polar codes) \cite{tal2015list} allows the list decoder to identify the correct path in the decoding list, significantly improving error correction performance.

Polarization-adjusted convolutional (PAC) codes \cite{arikan2019sequential} are a variant of polar codes \cite{arikan2009channel} resulting from the convolutional pre-transformation before polar coding. 
The pre-transformation in PAC coding can reduce the number of minimum weight codewords of underlying polar codes due to the impact on the formation of minimum weight codewords \cite{rowshan2021convolutional} and the involvement of frozen coordinates carrying non-zero values. This reduction is expected to improve the performance of PAC codes under (near) ML decoders, such as the list decoder \cite{tal2015list, rowshan2020polarization, yao2021list} and sequential decoders \cite{rowshan2020polarization}. % and sphere decoder \cite{kahraman2012code}. 

The coset-wise study on the reduction of minimum weight codewords (MWCs) in PAC coding \cite{rowshan2023minimum} revealed that there are limitations to this reduction.
%Studies on precoding schemes have been limited. 
%
% In our previous work \cite{Xinyi_Selective}, convolutional precoding was applied in reverse order for PAC codes, resulting in a notable reduction in the number of minimum weight codewords for high-rate short codes. 
%
Various pre-transformations (or precoders) have been proposed in the literature, including those based on dynamic frozen bits, parity bits, CRC bits, and pre-transformations in PAC coding \cite{rowshan_convolutional_2021, liu2023novel, wan_enhanced_weight_2023, gu_2023_improved, Zunker_rowmerged_2023}.  Deep polar codes, constructed in series, were introduced in \cite{choi2024deep}, followed by sparsely pre-transformed polar codes, designed in parallel to reduce encoding and decoding complexity, in \cite{choi2024sparsely}. However, these approaches do not address precoding from the perspective of MWC formation. A detailed overview of polar codes, PAC codes, and their variations can be found in \cite[Section VII]{rowshan_OJCOMS_2024}.
% In \cite{Samir_selectiveky_2020}, it was suggested to limit the precoding to frozen bits. %although it did not improve the error correction performance significantly. 
% A short-long bit-range precoder with two shift registers was proposed \cite{rowshan2021precoding}, where the first shift register was the conventional PAC precoder and the second was used to combine a subset of previous bits. Precodings of pac codes considering the weight distribution were studied in \cite{Chen_weighted_pac_2023, Liu_hamming_check_2023, wan_enhanced_weight_2023}.
%

In \cite{gu2024reverse}, we introduced reverse PAC (RPAC) codes, which address one of the key conditions in the limitation of PAC codes by performing convolutional precoding in reverse order. This approach significantly reduces the number of MWCs for high-rate short codes, leading to notable improvements in error correction performance.

%In this work, we propose a scheme to overcome another condition in the limitation of PAC codes. By reserving large-index coordinates in the rate profile, this limitation can be addressed. We observed a similar phenomenon in CRC-polar codes and analyzed the formation of MWCs in CRC-polar codes, providing a method to enumerate them. 
In this work, we address another condition in the limitation of PAC codes by reserving large-index coordinates in the rate profile. Observing a similar phenomenon in CRC-polar codes, we analyze the formation of MWCs in CRC-polar codes and provide a method to enumerate them. 
Drawing from the reduction of MWCs in PAC and CRC-polar codes, we propose two novel schemes: 1. Profile-shifted PAC (PS-PAC) codes, which introduce no-freedom large-index coordinates in the rate profile of PAC codes to prevent MWC formation and reduce their number.
2. Continuous CRC-polar (CCRC-polar) codes, which replace frozen bits with remainders from continuous divisions of the partial message and the CRC polynomial.
Simulation results show that PS-PAC codes achieve up to a 0.5 dB power gain for short codes and 0.1–0.2 dB for long codes compared to PAC and CRC-polar codes. For long codes, CCRC-polar codes provide an additional 0.12 dB improvement over CRC-polar codes.
%Drawing insights from the reduction of MWCs in both PAC and CRC-polar codes, we propose two novel schemes. 
%The first, profile-shifted PAC (PS-PAC) codes, includes large-index coordinates in the rate profile of PAC codes, preventing the formation of MWCs and reducing their overall number. The second, continuous CRC-polar (CCRC-polar) codes, continuously precodes the frozen bit coordinates with the intermediate remainders from CRC checks. Simulation results demonstrate that PS-PAC codes achieve up to a 0.5 dB power gain for short codes compared to PAC and CRC-polar codes and 0.1-0.2 dB for long codes. CCRC-polar codes achieve an overall improvement of 0.1 dB compared to CRC-polar codes for long codes.

%%%%%%%%%%%%%%%%%%%%%%%%%%%%%%%%%%%%%%%%%%%%%%%%%%%%%%%%%%%%%%%%%%%%%%
\section{PRELIMINARIES}\label{sec:prelim}
%\subsection{Notations}
%In this section, we first introduce some notations, and then we briefly review polar and PAC coding schemes, maximum likelihood decoding for a binary input additive white Gaussian noise (BI-AWGN) channel as well as sphere decoding algorithm as a candidate decoder for the proposed precoding scheme. 
%We denote by $\ft$ the finite field with two elements. The cardinality of a set is denoted by $|\cdot|$.  %The interval $[a,b]$ represents the set of all integer numbers in $\{x:a\leq x\leq b\}$. %The \emph{support} of a vector $\be = (e_1,\ldots,e_{n}) \in \ft^n$ is the set of indices where $\be$ has a nonzero coordinate, i.e. $\supp(\be) \triangleq \{i \in [1,n] \colon e_i \neq 0\}$ . %The indices of the vector elements may start from 0 or 1. 
Notations: %We denote the \emph{weight} of a vector $\be \in \ft^n$ by $w(\be)\triangleq |\supp(\be)|$. The support $\supp(\be)$ is the set of indices where $\be$ has a nonzero coordinate.
We denote the set of indices where vector $\be \in \ft^n$ has a nonzero coordinate by support $\supp(\be)$. The \emph{weight} of $\be$ is $\w(\be)\triangleq |\supp(\be)|$. 
Let $[a,b] \triangleq \{a,a+1,\cdots, b\}$ denote a subsets of consecutive integers.
%The summation in $\ft$ is denoted by $\oplus$. Let $[\ell,u]$ denote the range $\{\ell,\ell+1,\ldots,u\}$ and bold letters denote vectors.
The (binary) representation of $i \in [0,2^n-1]$ in $\ft$ is defined as $\bin(i)=i_{n-1}...i_1i_0$, where $i_0$ is the least significant bit, that is $i = \sum_{a=0}^{n-1}i_a 2^a$. 
%A $K$-dimensional subspace $\cC$ of $\ft^N$ is called a linear $(N,K,d)$ \emph{code} over $\ft$ 
We use the operator $\backslash$ in $\mathcal{A}\backslash\mathcal{B}$ to subtract elements of the set $\mathcal{B}$ from $\mathcal{A}$.
The notation $v^{b}_{a}$ represents a vector sequence with the indices ranging from $a$ to $b$, i.e. $[ v_a, v_{a+1}, ..., v_{b-1}, v_{b} ]$.

\subsection{Polar codes, CRC-Polar Codes, and PAC Codes} %%%%%%%%%%%%%%%%%%%%%%%%%%%%%%%%%%%%%%
\label{subsec:RMPolar}
Polar codes of length $N=2^n$ are constructed based on the $n$-th Kronecker power of binary Walsh-Hadamard matrix  
$\mathbf{G}_2 = 
{\footnotesize \begin{bmatrix}
1 & 0 \\
1 & 1
\end{bmatrix} }$, that is, $\bGN=\mathbf{G}_2^{\otimes n}=[\bg_0\;\;\bg_2\;;\cdots\;\;\bg_{N-1}]^T$ which we call it {\em polar transform} throughout this paper. %We denote the element $j$ in row $\bg_i$ of the polar transform by $g_{i,j}, i,j\in[0,N-1]$. 
A generator matrix of the polar code is formed by selecting the rows $\bg_i,i\in\I$ of $\bGN$. Then, $\CI$ denotes such a linear code. Note that $\I \subseteq [0,N-1]=[0,2^n-1]$. The characterization of the information set $\I$ for polar codes is based on the channel polarization theorem \cite{arikan2009channel} and the concept of \textit{bit-channel reliability}. %A polar code of length $N=2^n$ is constructed by selecting a set $\I$ of indices $i\in[0, N-1]$ with high reliability \cite{arikan2009channel}. 
The indices in $\I$ are allocated for $K$ information bits with $[i_0,\; i_1\; ...\; i_{K-1}] \in \I$ representing the elements in $\I$. The indices in $\mathcal{I}^c \triangleq [0, N-1]\setminus \I$ are used to transmit a known value, `0' by default, which are called \emph{frozen bits} and the corresponding rows are frozen rows. 
%Regardless of the method, we use for forming the set $\mathcal{I}$ for a polar code, the bit-channels with indices in the set $\mathcal{I}$ must be more reliable than any bit-channels in $\mathcal{I}^c$. %The notation $W^{(j)}_N\preceq W^{(i)}_N$ is used to say that the bit-channel $i$ is more reliable than bit-channel $j$. 

Polar codes are precoded with CRC codes to enhance error correction performance \cite{tal2015list}. 
In this work, we employ systematic CRC codes with polar codes, referring to the resulting precoded codes as CRC-polar codes. %, appending the CRC bits at the end of the information sequence by default. 
Let $t$ denote the degree of the CRC polynomial and let $\bq = [q_t, q_{t-1}, \cdots, q_0]$ represent the coefficient vector for the CRC generator polynomial. %For CRC-polar codes, CRC bits are appended before the polar transformation. 
The generator polynomial for these CRC codes can be expressed as $q(x) = q_t x^{t} + q_{t-1} x^{t-1} + \cdots + q_0$. 
The generator matrix of the CRC codes, denoted as $\bG_c$, can be constructed by dividing each row of an identity matrix $\bI$ with size $K\times K$ by the CRC polynomial $q(x)$ and taking the remainder:
\begin{equation} \label{eq:G_crc}
    \bG_{c} = [\,\bI\; | \; \text{remainder}(\, \tfrac{\bI}{q(x)} \,) \,].
\end{equation}
For a data sequence $\bd$, the precoded vector $\bc$ for CRC-polar codes can be obtained either using the generator matrix as $\bc = \bd\,\bG_c$, or equivalently by dividing $\bd$ by $q(x)$:
% remainder $d(x)$
\begin{equation} \label{eq:CRC_encode}
    \bc = [\, \bd\; |\; \text{remainder} (\, \tfrac{\bd}{q(x)} \,].
\end{equation}
%We use $\bR = [R_0,\, R_1,\, \cdots,\, R_t]$ to refer to the remainder sequence of the CRC coding. 
The input vector to the polar transformation, denoted as $\mathbf{u}=[u_0,\ldots,u_{N-1}]$, is obtained through rate profiling based on the selected information set $\I$ for the vector $\bc$. We denote the set of bit coordinates assigned to the CRC bits as $\R$. %, where $\R = [N-t,\, N-1]$. 
The resulting $\mathbf{u}$ is mapped to codeword $\mathbf{x} = \mathbf{u}\mathbf{G}_N$ via the polar transformation. 
%After rate profiling based on the selected information set $\I$ for the vector $\bc$, The input vector $\bu$ is obtained by the 
%rate profile $\bc$, the vector $\mathbf{u}$ is mapped to codeword $\mathbf{x}=\mathbf{u}\mathbf{G}_N$.
 
In polarization-adjusted convolutional (PAC) coding \cite{arikan2019sequential}, a pre-transformation stage is introduced between the rate profiling and polar coding stages. During this stage, the input vector $\bu$ for polar coding is obtained through a convolutional transformation using the binary generator polynomial of degree $s$, with coefficients $\mathbf{p}=[p_0,p_1\ldots,p_s]$ as follows: 
\begin{equation}\label{eq:pac_precoding}
    u_i = \sum_{\ell=0}^s p_\ell v_{i-\ell},
\end{equation} 
where $\mathbf{v}=[v_0,\ldots,v_{N-1}]$ is the vector constructed based on $\I$. 
%This convolutional transformation combines $s$ previous input bits with the current input bit $v_i$ to calculate $u_i$.
%This coding scheme is called {\em polarization-adjusted convolutional (PAC)} coding. %The parameter $s$ is known as the {\em memory} of the shift register and by including the current input bit  we have the {\em constraint length} $s+1$ of the convolutional code. 
The convolution operation can be represented in the form of an upper triangular matrix \cite{rowshan_convolutional_2021} where the rows of the {\em pre-transformation matrix} $\mathbf{P}$ are formed by shifting the vector one element at a row. 
Note that $p_0=p_s=1$ by convention. Then, we can obtain $\mathbf{u}$ by matrix multiplication as $\mathbf{u}=\mathbf{v}\mathbf{P}$.
Due to this precoding, we would have $u_i\in\{0,1\}$ for $i\in\mathcal{I}^c$, indicating that $u_i$ corresponding to a frozen $v_i=0,i\in\I^c$ may no longer be fixed. 
Overall, we obtain $\mathbf{x}=\mathbf{v}\mathbf{P}\mathbf{G}_N$. %The concept of dynamic frozen bits for polar codes was initially introduced in \cite{Trifonov_2013_Dynamic}.

The code rates $R$ of the underlying polar codes, CRC-polar codes, and PAC codes are defined as the ratio of the data sequence length to the code length, expressed as $R = \frac{K}{N}$.

\begin{figure}
    \vspace{-5pt}
    \centering
    \includegraphics[width=1\columnwidth]{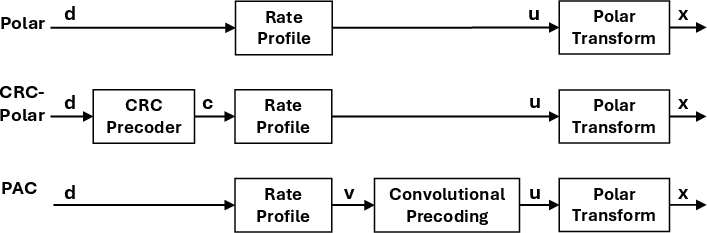}
    \caption{Precoding and encoding of polar codes.}
    \label{fig:encoding_polar_CRC_PAC}
    %\vspace{-10pt}
    \vspace{-8mm}
\end{figure}

\subsection{Minimum Weight Codewords in Cosets}%the Proposed precoding} %%%%%%%%%%%%%%%%%%%%%%%%%%%%%%%%%%%%
%\mohammad{We should remove the frozen coset term. }

It was analytically shown in \cite{rowshan_convolutional_2021, rowshan2023formation} that by convolutional pre-transformation, the number of minimum weight codewords, a.k.a error coefficient which is denoted by $\Awm$ where $\wm$ is the minimum weight, may significantly decrease relative to polar codes (without pre-transformation). Hence, from the union bound \cite[Sect. 10.1]{lin_2004_error}, we expect that this reduction potentially improves the block error rate (BLER) of a (near) maximum likelihood decoding for a binary input additive white Gaussian noise (BI-AWGN) channel. %, particularly at high signal-to-noise ratios (SNRs). 

In the conventional PAC coding, forward convolution as per \eqref{eq:pac_precoding} is performed. Although forward convolution can reduce the number of codewords of minimum weight relative to polar codes \cite{rowshan_convolutional_2021}, it has its own limitations. To show the limitations, we first partition all codewords, excluding the all-zero codeword, of a polar code $\C(\I)$ into \emph{cosets} defined as: %follows \cite[Defenition 3]{rowshan2023formtion}:
\begin{definition} \label{def:coset}
    %(\cite{rowshan-pac1}) 
    Cosets: Given information set $\I \subseteq [0,N-1]$ for a polar code, we define the set of codewords $\CiI\subseteq \CI$ for each $i\in \I$ in a coset of the subcode $\C(\I \setminus [0,i])$ of $\C(\I)$~as 
    \begin{equation}\label{eq:Ci}
        \CiI \triangleq \left\{\bgi+\sum_{h\in \H} \bgh \colon \H \subseteq \I \setminus [0,i]\right\}\subseteq \C(\I),
    \end{equation}
    where $\bgi$ is the \emph{coset leader}. We denote the number of minimum weight codewords of the coset $\Ci$ by $\Aiwm$. The total number of minimum weight codewords for a polar code $\C(\I)$ is $\Awm=\sum_{i\in\I}\Aiwm$.
\end{definition}

Observe that the coordinate of the first non-zero element in vector $\bu$, $i=\min\{\supp(\bu)\}$, while encoding by $\mathbf{x}=\mathbf{u}\mathbf{G}_N$, %plays a key role in classifying the codewords into cosets. %According to \cite[Lemma 1]{rowshan2023minimum}, this coordinate remains the same after precoding by $\bu=\bv\mathbf{P}$. 
% \end{equation}
%Nevertheless, t
the resulting $u_j$ for $j>i,j\in\I^c$ might be $u_j\not=0$, unlike in polar coding. %Hence, although due to retaining $i=\min\{\supp(\bu)\}$ in PAC coding, the minimum distance remains the same as polar codes, that is,
% \begin{equation}
%     d_{min} = w_{min} = \min(\{w(\bgi):i\in\I\}),
% \end{equation}
This difference may impact the number of minimum weight codewords in the cosets due to the inclusion of rows $\bg_j$ for $j\in\I^c\cap[i,N-1]$ in row combinations. %Note that the weight of the codewords in a coset $\Ci$ depends on the set of row indices $\H=\{h:h>i\}$ in \eqref{eq:Ci}. 
%
% Furthermore, according to \cite[]{rowshan2023formation}, 
% % \begin{equation}\label{eq:geq_wi}
% %     \w(\bgi+\sum_{h\in\mathcal{H}}\mathbf{g}_h)\geq \w(\bgi),
% % \end{equation}
% there exists minimum weight codeword $\bc\in\Ci$, where $\w(\bc)=\wm$, only if $i\in\mathcal{B}$ where set $\mathcal{B}$ is defined as
% \begin{equation}
%     \mathcal{B}=\{j:j\in\I,w(\bg_j)=\wm\}.
% \end{equation} 
Observe that we have $\H\subseteq[i+1,N-1]$ in PAC coding whereas in polar coding, we have $\H\subseteq[i+1,N-1]\backslash\I^c$. %Removing the constraint on $\H$ and considering all rows may give different codewords.  
The minimum distance of the PAC codes is \cite[Lemma 1]{rowshan2023minimum} 
% \begin{equation}
    $d_{\min} = \wm = \min(\{\w(\bgi):i\in\I\}).$
% \end{equation}

% Now, as the minimum distance of a polar code equals the minimum weight of the non-frozen rows in the polar transform, that is, 
% \begin{equation}
%     d_{min} = w_{min} = \min(\{w(\bgi):i\in\I\}).
% \end{equation}

% On the other hand, according to Corollary 5 in \cite{rowshan2023formtion}, we~have
% \begin{equation}\label{eq:geq_wi}
%     w(\bgi\oplus\bigoplus_{j\in\mathcal{H}}\mathbf{g}_h)\geq w(\bgi),
% \end{equation}
% where $\mathcal{H}\subseteq [i+1,2^{n}-1]$. Hence, the number of the minimum weight codewords denoted by $\Aiwm$, can be formed in the cosets $\CiI$ where the coset leader has weight $w(\bgi)=w_{min}$.

% In the forward convolution used in the conventional PAC codes, since for every $\CiI$, we have $i\in\I$, the minimum distance of polar codes is preserved. 

% Observe that in polar coding, we always have $b=0$ at coordinates $i\in\cI^c$ whereas $b\in\{0,1\}$ in PAC coding depending on the choice of coefficient vector $\bp$. 
%\textcolor{blue}{
According to \cite[Theorem 1]{rowshan2023formation}, the minimum weight codewords are uniquely formed by the following row combinations:
\begin{equation}\label{eq:wt_gi_gj_gm}
    \w\big(\bgi+\sum_{j\in\J}\bgj +\sum_{m\in\M(\J)}\bgm\big) =  \wm,
\end{equation}
% \begin{equation}\label{eq:wt_gi_gj_gm}
%     w\big(\bgi\oplus\bigoplus_{j\in\J}\bgj \oplus \bigoplus_{m\in\M(\J)}\bgm\big) =  w_{min},
% \end{equation} 
%where $\bgi$ is the leading row, $\bg_j, j\in\J$ are the core rows, and $\bg_m, m\in\M(\J)$ are the balancing rows. 
where $\w(\bg_i)=\wm$, $\J\subseteq\Ki$ and $\Ki$ is \cite[Lemma 2.a]{rowshan2023formation}
\begin{equation}\label{eq:set_Ki}
\Ki \triangleq \{j \in \I\backslash[0,i]\colon |\supp(j)\backslash\supp(i)|=1\}.
\end{equation}
As a result, every subset of $\Ki$ along with other rows in \eqref{eq:wt_gi_gj_gm} form a minimum weight codeword. 
The number of subsets of $\Ki$ is given by $2^{|\Ki|}$. Given $\B\triangleq\{i\in\I:\w(\bgi)=\wm\}$, the total number of minimum-weight codewords of the polar code will be $\sum_{i\in\B}2^{|\Ki|}$. 
%
%Since we can have $2^{|\Ki|}$ subsets of any set in total, we can have this many minimum-weight codewords. 
The set $\M(\J)$ is a function of the set $\J$ and every $m\in \M(\J)$ has the property (see \cite[(9),(10)]{rowshan2023formation} for a detailed definition of $\M(\J)$):
\begin{equation}\label{eq:set_M}
\M(\J)\!\subseteq\!\{m\!>\!i:|\supp(\bin(m))\backslash\supp(\bin(i))|\!>\!1\}.
\end{equation} 
%
% The number of minimum weight codewords that are generated by the leading row $\bgi$ is denoted by $A_{i, \wm}$.
% where the general properties of sets $\J$ and $\M$ are 
% \begin{equation}\label{eq:set_J}
% \J\!\subseteq\!\{j\!>\!i\!:\!|\supp(\bin(j))\backslash\supp(\bin(i))|\!=\!1\},
% \end{equation} 
% According to \cite{rowshan2023formtion} and \cite{rowshan-pac_enum}, the minimum weight codewords are uniquely formed by row combinations in matrix $\bG_N$ as
% \begin{equation}\label{eq:wt_gi_gj_gm}
%     w\big(\bgi\oplus\bigoplus_{j\in\J}\bgj \oplus \bigoplus_{m\in\M(\J)}\bgm\big) =  w_{min},
% \end{equation} 
% where the general properties of sets $\J$ and $\M$ are 
% \begin{equation}\label{eq:set_J}
% \J\!\subseteq\!\{j\!>\!i\!:\!|\supp(\bin(j))\backslash\supp(\bin(i))|\!=\!1\},
% \end{equation} 
% \begin{equation}\label{eq:set_M}
% \M\!\subseteq\!\{m\!>\!i:|\supp(\bin(m))\backslash\supp(\bin(i))|\!>\!1\}.
% \end{equation} 
%Note that for every set $\J$, there exists a unique set $\M$ with the general property in \eqref{eq:set_M}. In this paper, we do not need to know how set $\M$  is formed as a function of $\J$ and the difference between these two sets suffices for our discussions. %Note that the elements of set $\J$ and $\M$ are in $(i,N-1]$.
The relation \eqref{eq:wt_gi_gj_gm} can be extened such that the sets $\J$ and $\M$ also intersect with $\I^c$ (see \cite[(18)]{rowshan2023minimum}). This is useful when considering the impact of precoding.
% \begin{remark}\label{rem:w_min_sabotage}
%     According \eqref{eq:wt_gi_gj_gm},  observe that if for some $\J$, we do not have the corresponding set $\M(\J)$  (although we might have $\M\not=\emptyset$), the generation of minimum weight codewords in coset $\Ci$ can be avoided. Conversely, this is the case if, for some $\M$, there is no corresponding $\J$ (including the case $\J=\emptyset$). This frequently happens as a result of precoding. %where the control over the value of $u_i,i\in\I^c$ is limited in the mapping $v_i\rightarrow u_i$ knowing $v_i=0$.
% \end{remark}
%}
Now, let us see the main limitation of the forward convolution in PAC coding, %According to \cite[Remark 2]{rowshan-pac_enum}, there are two categories of cosets %characterized based on any $j\in\I^c$ for $j>i$ 
that forward convolution cannot reduce $\Aiwm$.  
\begin{lemma}\label{lem:incapble}
    (\cite[Lemma 2]{rowshan2023minimum}) For any coset $\CiI$ where
    \begin{enumerate}
        \item $\I^c \cap(i, N-1]=\emptyset$, or
        \item $|\supp(\bin(f)) \!\backslash\! \supp(\bin(i))|\!=\!1, \forall f\!\in\!\left(\I^c\!\cap\!(i,N\!-\!1]\right)$,
    \end{enumerate}
    we have
    \vspace{-5pt}
    $$
    A_{i, \wm}(\bG, \I)=A_{i, \wm}(\mathbf{P G}, \I) .
    $$
    
    In other words, any cosets $\mathcal{C}_i$ where there is no frozen row $\mathbf{g}_f$ for $f \in \mathcal{I}^c \cap(i, N-1]$ such that $|\supp(\bin(f)) \backslash \supp(\bin(i))|>1$, we get $A_{i, \wm}=$ $2^{\left|\mathcal{K}_i\right|}$ in the PAC coding, independently of the choice of $\mathbf{p}$.
    \end{lemma}
%The obvious category is where we have $\H\cap\I^c=\emptyset$ for a coset, i.e., there exists no row $\bg_f$ for $f\in\I^c$ in the coset. The second category is where $\H\cap\I^c\not=\emptyset$ but for every $f\in\H\cap\I^c$, we have $|\supp(\bin(f))\backslash\supp(\bin(i))|=1$. 
%Hence, there is a lower bound for the error coefficient in the conventional PAC coding with forward convolution. 

%In the next section, we propose a different precoder that overcomes the limitations of the forward convolution.
% \begin{remark}
%     As forward precoding regardless of the choice of polynomial $\bp$ is not effective on some cosets characterized in \cite[Remark 2]{rowshan-pac_enum}, the capability of reducing the error coefficient of polar codes by forward precoding is limited.
% \end{remark}

\subsection{Error Occurrence in List Decoding}\label{ssec:gen_approach} 
% types of error, relation with path metric

List decoding for polar codes and CRC-polar codes involves two types of errors, as analyzed in \cite{wilson2021error}. Type I errors ($\EI$) occur when the correct codeword is not included in the decoding list, typically due to an insufficient list size $L$ or a low signal-to-noise ratio (SNR). These errors are common to both polar and CRC-polar codes. % and depend solely on the list formation process. 
Type II errors ($\EII$) arise when the correct codeword exists in the final decoding list but %with %less likelihood and 
is not selected as the output of the list decoder. For polar codes, $\EII$ occurs due to small minimum Hamming distances or poor likelihood differentiation. For CRC-polar codes, $\EII$ occurs when an incorrect codeword passes the CRC check and is selected. Type II errors decrease with longer CRC lengths $t$ or stronger CRC polynomials. %The probability of $\EII$ for list decoding of CRC-polar codes is formulated in \cite[(17)]{wilson2021error}.

% \begin{equation} \label{eq:E_II_prob}
%     P(\EII) \simeq \frac{L-1}{2^{-t}} Q\left(\left[\left(2 E_b / N_0\right)\left(\dm R\right)\right]^{1 / 2}\right).
% \end{equation}
%saturates in $L$ and eventually dominates the overall BLER
%Thus, increasing $t$ or increasing $\dm$ is expected to reduce the probability of this type of error. 

%an incorrect codeword is selected despite the correct codeword being present in the list.

%
%The list decoders have type I errors denoted as $E_I$ when the list decoder misses the correct codeword in the decoding list. The probability of type I errors depends on the SNRs and the list sizes $L$ of the list decoder. This type of errors occurs when the list size is insufficient or the signal-to-noise ratio (SNR) is low. Error $E_I$ is common to both polar codes and CRC-polar codes, as it depends solely on the list formation process.

%%%%%%%%%%%%%%%%%%%%%%%%%%%%%%%%%%%%%%%%%%%%%%%
\section{Reserving Large-index Coordinates in Rate-profile to Enhance Weight Distribution}\label{sec:fz_indices} 
% No information transmission via these coordinates

To address the limitations of incapable cosets described in Lemma \ref{lem:incapble} and further reduce the total number of minimum weight codewords $A_{\wm}$, we proposed a solution in our previous work \cite{gu2024reverse}, named RPAC codes, which removes the second condition of Lemma \ref{lem:incapble}.
In this work, we focus on addressing the first condition. 
The idea is to incorporate more frozen bits at large-index coordinates in the rate profile, ensuring that $\I^c \cap(i, N{-}1]\neq \emptyset$, thereby violating Lemma \ref{lem:incapble}.1. 

To further refine the problem, we aim to reduce $\Awm$ in PAC codes by introducing frozen coordinates $f$ in the rate profile, satisfying the following conditions:
\begin{enumerate}
    \item $f \in \I^c \cap(i, N{-}1]$. These coordinates belong to sets $\J$ and $\M(\J)$ of $\Ki$ in a specific $\CiI$,  i.e. $f \in \J \cup \M(\J)$, with $ \J\subseteq \Ki$ and $\w(\bgi) {=} \wm$, which contribute to forming MWCs in polar codes, as described in \eqref{eq:wt_gi_gj_gm}, \eqref{eq:set_Ki} and \eqref{eq:set_M}.
%$f \in \J \cup \M(\J)$, where 
%Reserving these large-index coordinates as frozen in the rate profile for PAC codes,   

    \item $[f-s,\, f)\cap \I \neq \emptyset$, ensuring $u_f$ is precoded with \eqref{eq:pac_precoding}. These bits, instead of being frozen, become no-freedom bits, restricting the flexibility to assign $u_f = 0$ or $u_f = 1$.

    %     \item $f \in \J \cup \M(\J)$, where $ \J\subseteq \Ki$ and $\w(\bgi) = \wm$. 
    %     % where $\J$ and $\M(\J)$ are essential sets of a specific coset $\CiI$ that contribute to forming MWCs in polar codes.
    %     \item $[f-s+1,\, f]\cap \I \neq \emptyset$. 
\end{enumerate}
% Reserving these large-index coordinates as frozen in the rate profile ensures that, for PAC codes, they are precoded with the forward convolution and lack the freedom to assign arbitrary $0$ or $1$. 

%this approach introduces the no-freedom set that have intersections between the sets $\J$ and $\M(\J)$ for $\CiI$ which forms MWCs in polar codes. 
%intersections between the no-freedom set $\F$ with the sets $\J$ and $\M(\J)$ 

\begin{remark} \label{remark:elimintae_MWC}
    To achieve both $f \in \I^c \cap(i, N{-}1] \neq \emptyset$ and $[f{-}s,\, f)\cap \I \neq \emptyset$, large-index coordinates need to be reserved in the rate profile. This adjustment could impact the minimum weight codewords in the coset $\Ci$ as follows:
    $$\w\big(\bgi+\sum_{j\in\J}\bgj +\sum_{m\in\M(\J) }\bgm +\sum\bgf \big) \neq \wm.$$
    % \begin{enumerate}
    % \item $\w\big(\bgi+\sum_{j\in\J}\bgj +\sum_{m\in\M(\J) }\bgm +\sum\bgf \big) \neq \wm$.
    % % \item 
    % \end{enumerate}
    The constraints introduced by $f$ could disrupt the row combinations necessary for forming MWCs, as described in \eqref{eq:wt_gi_gj_gm}. %, effectively reducing the total number of MWCs and enhancing the overall error correction performance of the codes. 
    By violating the conditions in \eqref{eq:wt_gi_gj_gm}, the MWCs are eliminated, leading to a reduction in $\Awm$. %based on $A_{i, \wm}(\mathbf{P}_N\bG_N, \I)$.
\end{remark}

% mention at the end of this section that This is happening in CRC-polar in serial concatenation. Then, you move to discuss it in detail in the next section. This idea was not intentional behind the CRC-polar or any other concatenated polar codes, but due to the needs for more sub-channels for the parities of the outer code, the large-index sub-channels played this role. Note that we recently discovered how the minium-weight codewords are formed in polar coding. 

This phenomenon occurs in CRC-polar codes through serial concatenation. Although this was not an intentional design choice for CRC-polar or other concatenated polar codes, the need for additional subchannels to accommodate the parities of the outer code naturally led to the large-index subchannels fulfilling this role. It is worth noting that the formation of minimum-weight codewords in polar coding has only been recently understood.

%%%%%%%%%%%%%%%%%%%%%%%%%%%%%%%%%%%%%%%%%%%%%%%
\section{Reduction of the number of MWCs % $\Awm$
in CRC-polar Codes}\label{sec:fz_indices} 
% discussing differences of polar,pac, and crc-polar codes: appending parties
% Different rate profiles utilize distinct subchannels of polar codes, resulting in significant variations in error correction performance. When precoders are concatenated with polar codes, the adjusted frozen bit coordinates may deviate from the reliability order.
%
In this section, we present an approach to analyze and enumerate the number of minimum weight codewords $\Awm$ in CRC-polar codes. %By considering the frozen bit coordinates, we analyze and compare polar codes, CRC-polar codes, and PAC codes in terms of the number of MWCs and their formation.

For polar codes, we have the information bits $u_i \in \{0, 1 \}, i\in \I$. 
In CRC-polar codes, the CRC bits are appended to the end of the data sequence. After the polar transform, these CRC bits occupy the $t$ most reliable bit coordinates of the polar codes, corresponding to the largest-index coordinates in the rate profile. %, corresponding to the last $t$ bit coordinates of the vector $\bu$. 
%
%hese bit coordinates are constrained for parity checks, leaving no freedom to assign arbitrary information values $0$ or $1$. %where $i \in \R$. 
Examining the CRC generator matrix along with the polar transformation, it can be observed that  CRC bits %$\bR$ in CRC-polar codes 
perform parity checks on multiple information bit coordinates. Specifically, $u_r = u_j \oplus u_k\oplus \cdots \oplus u_h$, where $r\in \R $ and $ j,k,h \in \I$. Thus, the values of $u_r$ are determined by the values of the corresponding information bit coordinates, lacking the freedom to independently assign information $u_r = 0$ or $u_r = 1$.

As shown in \eqref{eq:wt_gi_gj_gm}, for a certain coset $\CiI$, we need to have specific row combinations to form MWCs in polar codes. Consequently, if the CRC bits, derived from parity checks, fail to satisfy these conditions, %as defined in \eqref{eq:wt_gi_gj_gm}, 
the corresponding MWCs are effectively eliminated. 
%It can be observed from the CRC generator matrix that 
% + a connection and a summary of the paragraph 
Let us see an example to illustrate the elimination of WMCs in CRC-polar codes:

\begin{example} \label{exp:32_16_dmin}
    Consider a $(32,\, 16)$ polar code with $\I_{\text{Polar}} = \{11,13\!-\!15,19,21\!-\!31\}$, where `$-$' indicates a range of integers. This polar code has $\wm = 4$. For a CRC-polar code with the same rate and generator polynomial $q(x) = x^5+ x^3 +1$, the information set becomes $\I_{\text{CRC}} = \{7,11\!-\!15,17\!-\!31\}$. 
    % \begin{align*}
    %     \text{Polar:}  &\w(\bg_{12}+ \bg_{13}+\bg_{28}+ \bg_{29}) =4, \\
    %     \text{CRC-Polar:}  &\w(\bg_{12}+ \bg_{13}+\bg_{31}) =28.
    % \end{align*} 
    % \vspace{-3pt}
    \begin{equation*}
        \text{Polar:}  \w(\bg_{24}+ \bg_{25}) =4, 
        \vspace{-5pt}
    \end{equation*}
    \begin{equation*}    
        \text{CRC-Polar:}  \w(\bg_{24}+ \bg_{25}+\bg_{27}+\bg_{31}) =20.      
    \end{equation*}
    % \vspace{-2pt}
    With leading row $\bg_{24}$, a MWC of polar code can be formed if $\J= \{25\}$. However, in the CRC-polar code, the CRC bits occupy the last $t= 5$ %positions of the 
    information bit coordinates, i.e. $\R = \{27\!-\!31\}$. %There is no freedom in the assignment of $u_{28}$ and $u_{29}$. 
    With the same $u_{24}$ and $u_{25} = 1$, the parity checks at the CRC bit coordinates result in a fixed sequence that $\bu_{27}^{31}= [1,\, 0,\, 0,\, 0,\, 1]$. This leads to a different row combination, forming a higher-weight codeword with $\w = 20$, effectively eliminating the MWC of the original polar code.
    
\end{example}
%For instance, 
%Let $m_p$ be the set of the number of parity checked bits for each CRC bit. 
%The bit coordinates involved in the parity checks can be represented by the set $\bD = \{\D_t,\, \D_{t-1},\, ...,\, \D_0 \}$. For $0\le i \le t$, set $\D_i = \bg_{j} \oplus \bg_{} \cdots \oplus \bg_{}$ represent the parity checked by each CRC bit. 

%For $0\le i \le t-1$, the bit coordinates involved in the $i^{\text{th}}$ parity check bit can be represented by the set $\D_i = \bg_{j} \oplus \cdots \oplus \bg_{k}$, where $\bg_{j}$ and $\bg_{k},\, \{j,\,k\} \in [0, N-t],$ denote the rows in the polar transform matrix corresponding to the relevant information bit coordinates. 

% + Venn diagram, showing the relation between the sets and $\R$ 
As shown in the example, the coordinates $\R$ of CRC bits in CRC-polar codes may intersect with the sets $\J$ or set $\M(J)$ of  $\Ki$ in a specific coset $\CiI$, which are essential sets for forming MWCs in polar codes.
By analyzing $\R \cap \{\J \cup \M(\J)\}$ in different cosets, %the intersection of the sets $\J$ and $\M(J)$ with the set $\R$ of CRC-polar codes, i.e. $\R \cap \{\J \cup \M(\J)\}$, %considering the coordinates $[0,\, N-t]$, 
%where $\J$ and $\M(\J)$ are essential sets 
two conditions emerge under which the MWCs of polar codes can be eliminated, as descibed in Remark \ref{remark:elimintae_MWC}. For any $r\in \R$, with leading row $\w(\bgi) = \wm$ and $\J \subseteq \Ki$ the MWC of polar codes is canceled if:
% \vspace{-2pt}
\begin{equation} \label{eq:CRC_cancel_requirement}
    u_r = 
    \begin{dcases*}
        1, & if $r \notin \J \cup \M(\J)$,\\
        0, & if $r \in \J \cup \M(\J)$. \\
    \end{dcases*}
\end{equation} %, with the help of the appended $\R$: 
% For a MWC of polar codes, , under two conditions where concatenating CRC codes can avoid forming the MWCs:

By examining whether the appended CRC bits cancel the MWCs of polar codes individually, the number of MWCs in CRC-polar codes can be enumerated. This reduction in the total number of MWCs can enhance the error-correcting performance of the CRC-polar codes. Furthermore, if all codewords with $\wm$ in polar codes are eliminated due to the constraints imposed by the last $t$ CRC bits, the minimum distance $\dm$ of the resulting CRC-polar codes increase. %If all MWCs of polar codes with weight $\wm$ are cancelled out by the CRC bits, %are eliminated by the CRC bits, 
%the minimum distance of the CRC-polar codes increases, improving the error correction performance of the codes.

% \begin{enumerate}[1.]
%     \item $ \R \cap \{\J \cup \M(\J)\} = \emptyset$ and $u_{i} = 1,\, i\in \R$: 
%     When $u$ at coordinated $i \in [0,\, N-t]$ can form an MWC for polar codes, due to parity checks at the appended CRC bits $\R$, we may have $u_{\R} = 1$. Consequently, this could cause $I(\cO) = 0$ as indicated in \eqref{eq:I_O_indicator}. The occurrence of an undesired $u = 1$ during MWC formation can lead to a higher weight codeword, thereby decreasing the $\Awm$. 

%     \item $ \R \cap \{\J \cup \M(\J)\} \neq \emptyset$  and $u_{i} = 0,\, i\in \R \cap \{\J \cup \M(\J)\} $: When there is an intersection between the $\cO$ of polar codes and the CRC bit coordinates, forming MWCs requires setting $u_{\R}$ to 1. However, the values for CRC bits are obtained by parity checks. If the resulted $u_{\R} = 0$, the formation of MWC for the polar codes can be avoided. 
    
% \end{enumerate}

% PAC codes 

% \begin{equation}\label{eq:wt_decomp_PAC} 
%     \w \big( \bgi \oplus \underbrace{ \bigoplus_{j\in\J \cup{\F^{\J}} } {\bgj}}_{\text{core rows} } \oplus \underbrace{ \bigoplus_{ m\in \M(\J\cup{{ \F^{\J} })}} \bgm }_{\text{balancing rows}} \big) =  \dm,
% \end{equation} 

%%%%%%%%%%%%%%%%%%%%%%%%%%%%%%%%%%%%%%%%%%%%%%%
\section{Proposed Schemes}\label{sec:enhanced_pac}%padded 

% The forward convolutional precoding in PAC does not inherently provide error correction capability; however, it modifies the row combinations of the original polar codes, reducing the number of MWCs. The limitations of this reduction, particularly when the contribution of frozen rows to codeword formation is lacking, are highlighted in Lemma \ref{lem:incapble}. For incapable cosets, the pre-transformation cannot reduce the number of WMCs.
% % during decoding the frozen bit coordinates

% In contrast,  CRC-polar codes eliminate MWCs by having large-index and no-freedom bit coordinates in the rate profile, leveraging the intersection of CRC bit coordinates with the sets $\J$ and $\M(\J)$ for specific cosets that form MWC in polar codes, as described in \eqref{eq:CRC_cancel_requirement}. %However, the appended CRC bits occupy the $t$ most reliable subchannels of polar codes, which can pose challenges, particularly for short codes. 

% These precoders of polar codes alter row combinations and impact the formation of MWCs in distinct ways. The motivation lies in leveraging the mechanisms of MWC elimination in CRC-polar and PAC codes. By combining their respective approaches to MWC elimination, it is possible to harness the advantages of both precoders to improve performance. 
In the following subsections, we propose two schemes inspired by PAC codes and CRC-polar codes, focusing on their approaches to reducing MWCs.

%%%%%%%%%%%%%%%%%%%%%%%%%%%%%%%%%%%%%%%%%%%%%%%
\subsection{Profile-Shifted PAC Codes}\label{sec:enhanced_pac}%padded   Learning from Serial Concatenation in CRC-Polar Codes

%Different precoders of polar codes change the row combinations and affect the formation of MWCs in different ways. The motivation is learning from the elimination ways of MWCs from CRC-polar and PAC codes and combine the ways of their elimination of MWCS to gain the advantage from both precoders.

% Learning from CRC-polar codes, we propose introducing non-information bit coordinates in the last few bit coordinates of PAC codes. These coordinates lack the flexibility to assign $0$ or $1$. This could not only break the limitations of PAC codes but also introduce intersections with the sets $\J$ and $\M(\J)$ for $\CiI$ which form MWC in polar codes. By canceling out these MWCs, this method achieves a further reduction in $\Awm$ for PAC codes. We use Scheme I to refer to this scheme. 

% Let $\alpha$ denote the number of frozen bit coordinates placed at the end of the Scheme I codes. These frozen bit coordinates replace the $\alpha$ most reliable bit coordinates in the information set $\I$, similar to the approach in CRC-polar codes. The coordinates of these frozen bits are denoted by the set $\F$. %$[i_K-\alpha, i_K]$
% To maintain the same code rate, the $\alpha$ most reliable bit coordinates are removed from the frozen set and reassigned to $\I$ to carry information. 

Learning from serial concatenation in CRC-polar codes, we propose introducing non-information bits at large-index coordinates of PAC codes, replacing the most reliable bit coordinates in the information set $\I$. Let $\alpha$ denote the number of frozen bit coordinates placed at the end of rate-profile. These non-information bit coordinates are represented by set $\F$. To maintain the same code rate, the $\alpha$ most reliable bit coordinates are removed from the frozen set $\I^c$ and reassigned to $\I$ to carry information. We refer to this approach as profile-shifted PAC(PS-PAC) codes.

\begin{figure}[ht]
    \vspace{-5pt}
    \centering
    \includegraphics[width=1\columnwidth]{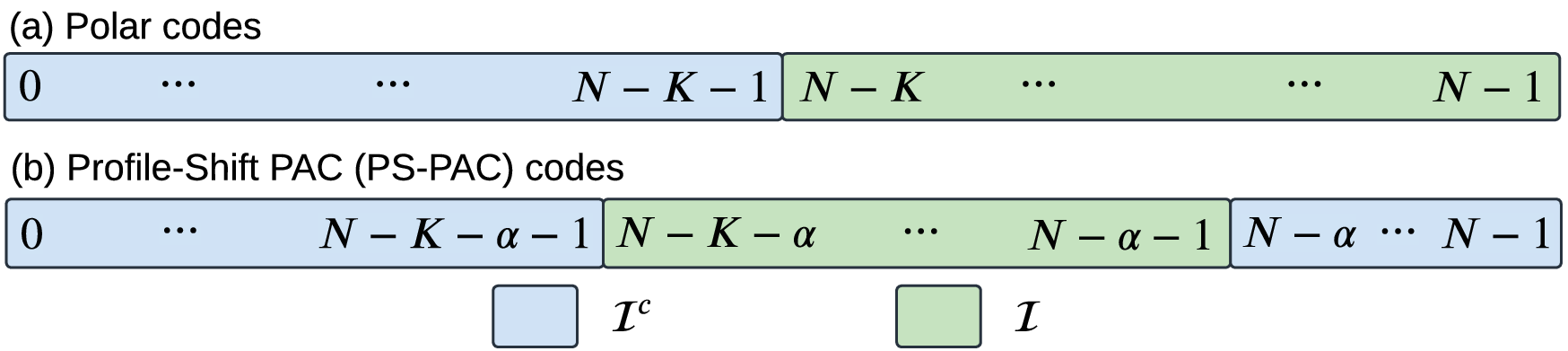}
    \caption{Constructions of polar and PS-PAC codes in reliability order.}
    \label{fig:PAC_SchemeI_rate_profile}
    \vspace{-10pt}
\end{figure}
Sequences of bit indices from $0$ to $N-1$ are depicted in Fig. \ref{fig:PAC_SchemeI_rate_profile}, illustrating the ordered subchannels of polar codes based on their reliability. The information and frozen sets of polar and PS-PAC codes are labeled, respectively, in Fig. \ref{fig:PAC_SchemeI_rate_profile}. 
The construction of PS-PAC codes can be expressed as 
\begin{equation} \label{eq:scheme_I_rate_profile}
    %\!\!\!\!\!\!\!\!
    u_i = 
    \begin{dcases*}
        \text{frozen}, & if $i \in $ [0$,\,$ N$\!-\!$K$\!-\!\alpha\!-\!1$] $\cup$ [N$\!-\!\alpha,\,$ N$\!-\!1$],\\
        \text{information}, & if $i \in $ [N$\!-\!$K$\!-\!\alpha,\,$ N$\!-\!\alpha\!-\!1]$. \\
    \end{dcases*}
    \vspace{-2pt}
\end{equation}
% The 
PS-PAC codes can be viewed as a modification of the rate profile of PAC codes, swapping the $\alpha$ most reliable bit coordinates between $\I^c$ and $\I$. This modification is guided by the requirements discussed in Section \ref{sec:fz_indices}. 
%the intersection of the no-freedom set $\F$ with the sets $\J$ and $\M(\J)$ of $\CiI$, which are responsible for forming MWCs in polar codes, as described in \eqref{eq:wt_gi_gj_gm}. 

The coordinates $f\in \F$ occupy the most reliable coordinates, %although they are frozen, 
located after the information coordinates. 
Provided a loose constraint $s > \alpha$, the condition $[f-s+1,\, f)\cap \I \neq \emptyset$ holds. These bits with coordinates $f\in \F$, precoded with forward convolution as described in \eqref{eq:pac_precoding}, act as parity check bits, leaving no freedom to assign $u_f = 0$ or $u_f = 1$. %(similar to the phenomenon in CRC-polar codes). 
Additionally, the no-freedom coordinates at the end of the rate profile lead to $\F \cap \{\J \cup \M(\J)\}$
%an intersection with sets $\J$ or $\M(J)$ 
of $\Ki$ in a specific coset. Phenomena similar to those in CRC-polar codes emerge. The MWCs can be eliminated if $u_f$ satisfies the conditions in \eqref{eq:CRC_cancel_requirement}, combating the row combinations of MWCs in \eqref{eq:wt_gi_gj_gm}. The limitations of PAC codes highlighted in Lemma \ref{lem:incapble}, thus, can be addressed, leading to a further reduction in $\Awm$ for PAC codes.

\subsection{Continuous CRC-Polar Codes}\label{sec:convolved_crcpolar} 

The frozen bit coordinates in the CRC-polar codes remain frozen throughout the precoding process. Inspired by the convolution in PAC coding, we propose replacing these frozen bits in CRC-polar coding with remainders derived from the partial message sequence and the CRC polynomial.
%we propose utilizing these frozen bit coordinates in CRC-polar codes by precoding them with the remainders derived from partial message sequence with the CRC polynomial. 
Unlike the standard CRC-polar codes, where CRC bits are appended only at the end of the data sequence, the proposed scheme continuously incorporates remainders generated during the computation of CRC bits (based on the partial data sequence). We refer to this approach as continuous CRC-polar (CCRC-polar) codes.

In CRC-polar codes, as described in \eqref{eq:CRC_encode}, the remainder is obtained by dividing $\bd$ with $q(x)$. The process starts by taking the first $t+1$ bits of $\bd$, where the coordinates of $\bd$ represented by $[i_0,\, \cdots,\, i_t]$ for $i_j\in\I,\, j\in[0,t]$, and dividing it by $q(x)$, yielding an intermediate remainder. The remainder is then continuously updated by incorporating the remaining bits of the data sequence, ultimately producing the final remainder used in standard CRC-polar codes. We denote these intermediate remainders as $\br$. % = [r_0,\, r_1,\, \cdots,\, r_t]$. 

\begin{figure}[ht]
    \vspace{-10pt}
    \centering
    \includegraphics[width=1\columnwidth]{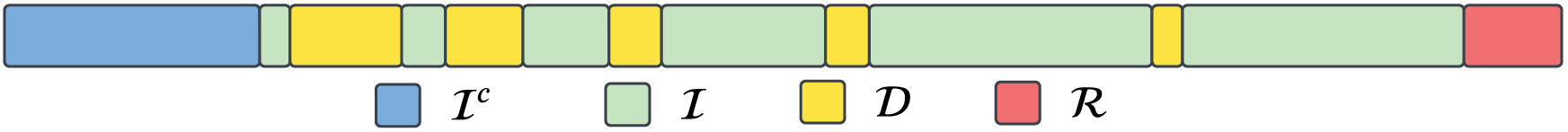}
    \caption{Construction of CCRC-polar codes in natural order.}
    \label{fig:CCRC_polar}
    \vspace{-10pt}
\end{figure}

The CCRC-polar codes leverage these intermediate remainders $\br$ by inserting them into the frozen bit coordinates as parity check bits, forming constraints. Since $\br$ is generated only after the CRC computation begins, the frozen bits located before the first information bits remain unchanged. %the frozen bit coordinates following the first information bit coordinate are precoded using these intermediate remainders. 
%Let $\D$ denote the set of frozen bit coordinates that are greater than the first element within 
Given $i_0=\min(\I)$, then we define %Specifically, 
% $\D = \{\beta:\beta > i_s,\; \beta\in\I^c\}$
\vspace{-8pt}
\begin{equation}
    \D = \{j: j > i_0,\; j \in\I^c\}.
\end{equation}
An example construction of CCRC-polar codes is %arranged in natural order and 
illustrated in Fig. \ref{fig:CCRC_polar}.

% for $ \in \I^c$ $u_i = \bR$ 

% 1. for  $i_0<i<i_s, i\in\D$,  for [$i_0, i_s$] frozen, filling the rest with all zero sequence 
% $\text{remainder} = \frac{[i, \bzero]}{q(x)}$ 

% 2. for  $i>i_s, i\in\D$,  for frozen bits after $i_s$, all the frozen bit coordinates are 
% $\text{remainder} = \frac{[i_j,i_k]}{q(x)}$

%During the precoding of CCRC-polar code, there are two conditions depending on the 
For coordinates $i_0\le i< i_t,\, i\in\I$,  the data sequence is insufficient to compute the first intermediate remainder. To resolve this, a zero sequence $\bzero$ is appended to the data sequence $[i_0, i]$ to reach the required length $t+1$ and $\br = [i_0,\, \cdots, i,\,\bzero]\, /\, {q(x)}$. For coordinates $i\ge i_t,\, i\in\I$, %intermediate remainder is computed as 
$\br = [i_{j-t},\, i_{j}]\, /\,{q(x)}$, where $t< j\le K$.

The bits at coordinates in $\D$ are replaced with $\br$ in a continuous manner. Let $i_f$ denote the first frozen coordinate after a block of information bits in the natural order of the rate profile (i.e. the first coordinate in each yellow block in Fig. \ref{fig:CCRC_polar}). 
Let $\beta_{i_f}$ represent the length of consecutive frozen coordinates (i.e. the length of coordinates in yellow blocks). %of CRC-polar codes. 
If %$\beta_{i_f}$ is shorter than the length of CRC remainder, i.e. 
$\beta_{i_f} \leq t$, then $\br$ is truncated to match the length of these frozen bits.  If $\beta_{i_f} >t$, $\br$ is cyclically repeated to fill all the frozen bits.
% 1. $\beta \leq s$
% 2. $\beta >s$
% frozen bit coordinates after
% if the length of consecutive frozen bit coordinates $\beta > s$, 

Different from the conventional list decoder for CRC-polar codes, which performs checksum verification at the end of the decoding, the list decoder for CCRC-polar codes performs CRC decoding bit by bit. To decode CCRC-polar codes, %it is essential to store and update the remainder for each path throughout the decoding process to place them at coordinates in $\D$ (similar to the encoding process) and use them for path metric computation. 
the remainder of each path must be stored and updated throughout the decoding process. These remainders are placed at the coordinates in $\D$ (as in the encoding process) and used for path metric computation.
%proceed with demapping over frozen coordinates. %to recover the frozen bits effectively. 
%When encountering frozen bit coordinates, the decoder 
%to places the corresponding remainder at the respective bit 79+4
%coordinate.

%%%%%%%%%%%%%%%%%%%%%%%%%%%%%%%%%%%%%%%%%%%%%%%
\section{Numerical Results}\label{sec:results} 

\begin{figure}[ht]
    \vspace{-10pt}
    \centering
    \includegraphics[width=0.9\columnwidth]{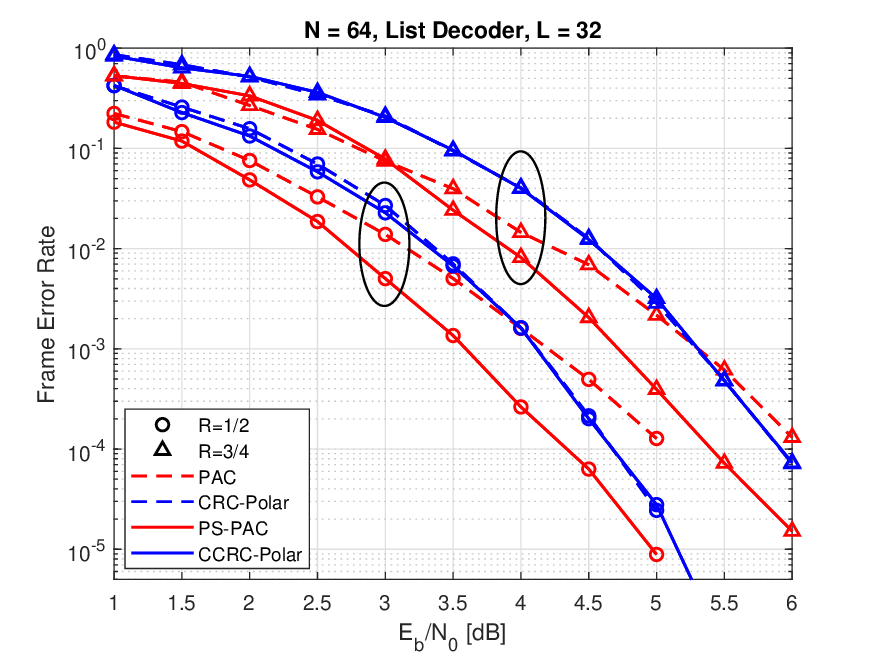}
    \caption{Performance of (64, 32) and (64, 48) codes.}
    \label{fig:bler_N_64}
    \vspace{-10pt}
\end{figure}

\begin{figure}[ht]
    \vspace{-10pt}
    \centering
    \includegraphics[width=1.0\columnwidth]{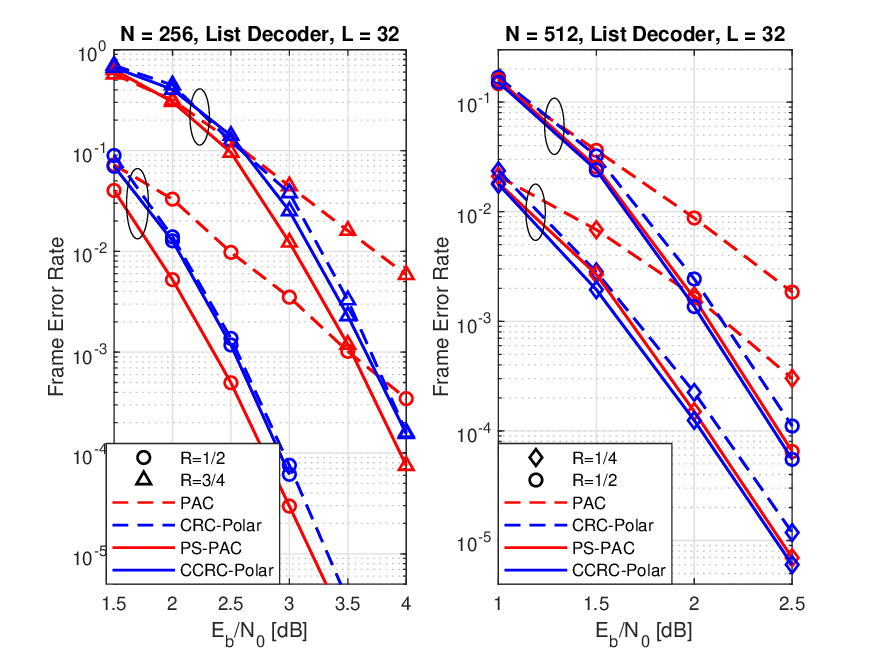}
    \caption{Performance of codes with $N = 256, \, 512$. 
    }
    \label{fig:bler_N_512}
    \vspace{-8pt}
\end{figure}

%In this section, we consider several example codes and numerically evaluate the proposed scheme in terms of error coefficient and block error rate followed by discussions. 
To demonstrate that the proposed schemes are effective for various code lengths and rates, the block error rates (BLER) of the codes with code lengths $N = 64,\, 256,\, 512$ are presented in Figs. \ref{fig:bler_N_64} and \ref{fig:bler_N_512}, constructed with approximate density evolution method \cite{Urbanke_Scaling_polar_2010}. %, while those of codes (512, 128), and (512, 256) are shown in Figs. 
%These codes are constructed using the approximate density evolution method \cite{Urbanke_Scaling_polar_2010}. %, Trifonov_efficient_2012}.  
% 
The error correction performance of the proposed PS-PAC codes and CCRC-polar codes are compared with that of PAC codes and CRC-polar codes under list decoding, using a fixed list size $L = 32$.
The polynomial $\bp = [1\;0\;1\;1\;0\;1\;1\;0\;1\;1]$ is employed for PAC and PS-PAC codes, with $\alpha {=} 8$ shifted bit coordinates in PS-PAC codes.
%The adopted CRC-polar codes use 11 CRC bits with the generator polynomial $g(x) {=} x^{11} +x^{10} +x^{9} +x^{5} +1$. 
The generator polynomial of the adopted CRC-polar codes is $q(x) = x^{11} +x^{10} +x^{9} +x^{5} +1$.
Table \ref{tb:Admin} provides the minimum distance with the corresponding error coefficient for short codes (64, 32) and (64, 48). %Note that concatenating these high-rate short codes with a short CRC can result in significant performance degradation due to a large rate loss. 

%A table to show the error coefficient of the codes before and after precoding, and after code modification.

% \begin{table}[ht] 
% % \vspace{-5pt}
% \setlength{\tabcolsep}{0.6em} % for the horizontal padding
% \renewcommand{\arraystretch}{1.2} %<- row spacing

% \caption{Minimum weight $\wm$ and the corresponding error coefficient $\Awm$ of polar codes ($^*$ refers to SR-PAC).}\label{tb:Admin}
% \centering
% \begin{tabular}{|l|l|l|l|l|}%llllllll} 
% \cline{1-5}
% {code} & {(64, 50)} & {(64, 32)} & {(64, 14)} & {(128, 110)} \\ 
% \cline{1-5}
% $\Awm$  & $A_{4}$           & $A_{8}$    & $A_{16}$   & $A_{4}$                \\  
% \cline{1-5}
% Polar                                        & 944   & 664   & 172     & 4099 \\

% \cline{1-5}
% SR/R-PAC(4)                                  & 435   & 339   & 220$^*$   & 1621 \\ 

% \cline{1-5}
% SR/R-PAC(7)                                  & 98   & 377$^*$  & 137$^*$    & 240  \\

% \cline{1-5}

% SR/R-PAC(10)                                 
% & 70$^*$     & 107$^*$       & 73$^*$                & 99$^*$                        \\
% \cline{1-5}
% \end{tabular}
% %\vspace{-5pt}
% \end{table}

% % table: 64,14, with Wmin
\begin{table}[ht] 
\vspace{-3pt}
\setlength{\tabcolsep}{0.6em} % for the horizontal padding
\renewcommand{\arraystretch}{1.2} %<- row spacing
\caption{Minimum weight $\wm$ and the corresponding error coefficient $\Awm$ of polar codes%($^*$ refers to SR-PAC)
.}
\label{tb:Admin}
\centering
\begin{tabular}{|l|l|l|l|l|}
\cline{1-5}
\multicolumn{1}{|c|}{\multirow{2}{*}{code}} & \multicolumn{2}{c|}{(64, 32)} %& \multicolumn{2}{c|}{(64, 14)} 
& \multicolumn{2}{c|}{(64, 48)} \\

\cline{2-5}
\multicolumn{1}{|c|}{}       & $\wm$ & $\Awm$                  %& $d_{min}$ & $\Adm$                  
& $\wm$ & $\Awm$           \\

\cline{1-5}
Polar                                          
& 8    & 664      %& 16   & 172     
& 4    & 432     \\

\cline{1-5}
PAC                                          
& 8    & 504      %& 16   & 140 
& 4    &  320    \\

\cline{1-5}
CRC-Polar                                   
& 8    & 6     %& 16   & 73        
& 4    & 13 \\ 
%\cline{1-7}
%FC-PAC(4)                                     
%& 4    & 435      & 16   & 220     & 4   & 1621      \\

% \cline{1-5}
% FC-PAC(7)                            that the     
% & 4    & 98      %& 16   & 137       
% & 4    & 240  \\

\cline{1-5}
PS-PAC                                  
& 8    & 27     %& 16   & 73        
& 4    & 27  \\

\cline{1-5}
CCRC-Polar                                   
& 8    & 63     %& 16   & 73        
& 4    & 306 \\ 

\cline{1-5}

\end{tabular}
 \vspace{-3pt}
\end{table}

In Fig. \ref{fig:bler_N_64}, with a short code length of $N = 64$, list decoding of CRC-polar code and CCRC-polar code show identical error correction performance for both moderate and high rate codes. This is due to the limited number of frozen bit coordinates available for precoding with the remainder for short codes, resulting in minimal impact on performance.
In contrast, the proposed PS-PAC codes outperform both PAC codes and CRC-polar codes for both (64, 32) and (64, 48) codes. As shown in Table \ref{tb:Admin}, about 94.64$\%$ of the minimum weight codewords of PS-PAC relative to PAC code are eliminated due to the introduction of large-index frozen bit coordinates, achieving up to a 0.5 dB improvement. While CRC-polar codes exhibit a steeper slope in high-SNR regimes, the power gain reduces as the SNR increases.

Relating performance to $\Awm$, both PS-PAC and CRC-polar codes significantly reduce $\Awm$ by incorporating no-freedom large-index coordinates. Additionally, forward convolution in PAC codes makes frozen bit coordinates dynamic, enhancing path metric penalties for incorrect paths during the decoding process \cite{Huazi_2018_PC_Polar}, effectively reducing type I errors $\EI$ \cite{wilson2021error}.
% CRC-polar codes and PAC codes reduce the number of MWCs and such that reduce the $\EI$ during the decoding processes.

% CRC increase $\dm$, which could reduce the type II errors as derived in \cite[(17)]{wilson2021error}

% \begin{figure}[ht]
%     \vspace{-5pt}
%     \centering
%     \includegraphics[width=0.9\columnwidth]{256_128_192_sum_results_isit_2025.eps}
%     \caption{Performance comparison of $(256,\, 128)$ and $(256,\, 192)$ codes. }
%     \label{fig:bler_N_512}
%     \vspace{-5pt}
% \end{figure}

%Similar conclusions can be drawn for longer codes at high rates, e.g., (128,110). As illustrated in Table \ref{tb:Admin} and Fig. \ref{fig:bler_(128,110)}, the considerable decrease in $\Awm$ for RPAC(10) enables the code to outperform its counterpart, PAC code. For CRC-Polar codes, the incorporated CRC polynomial introduces a rate loss in high-rate codes and involves very low-reliability bit-channels, thereby preventing the SCL decoder from significantly enhancing the error correction capabilities of polar codes. CRC-polar code with SCL decoder gives a sharp slope as the SNR increases, leading to a small gain between the proposed LA-SCL decoder for RPAC code. As a result, with $L=32$, about 0.1 dB improvement can be achieved when employing the LA-SCL decoder for the RPAC code. The LA-SCL(64) decoder for RPAC code demonstrates the same performance as the performance of SCL(128) decoder for the CRC-polar code and up to 0.2 dB power gain improvement can be achieved when the list size of LA-SCL decoder is increased to 128, bringing the curve closer to the performance of OSD for RPAC code.

In Fig. \ref{fig:bler_N_512}, similar trends are observed for longer codes in PS-PAC codes. 
For longer codes, PAC codes exhibit suboptimal performance. % as the forward convolution becomes less effective in reducing MWCs. 
However, by reserving large-index coordinates, PS-PAC codes achieve performance comparable to CRC-polar codes, resulting in an overall power gain of 0.1-2 dB.
For CCRC-polar codes, precoding on the frozen bit coordinates enhances the decoding ability to retain the correct path in the decoding list, allowing them to outperform CRC-polar codes and achieve an overall improvement of 0.12 dB when $N = 512$.

%As is shown in Fig. \ref{}, 

% \begin{figure}[ht]
%     % \vspace{-5pt}
%     \centering
%     \includegraphics[width=0.9\columnwidth]{64_14_sum_results_globecom.eps}
%     \caption{Performance comparison of (64,14) codes.}
%     \label{fig:bler_(64,14)}
%     % \vspace{-5pt}
% \end{figure}
%For the low-rate code (64,14), limited by a relatively small reduction in $\Awm$, the performance gain is not as significant as high-rate codes. As shown in Fig. \ref{fig:bler_(64,14)}, the low-rate CRC-polar code under the SCL decoder approaches the performance of the FC-PAC code as the SNR increases. Compared to low-rate codes under SCL and sphere decoding schemes, FC-PAC code can provide about 0.2 dB power gain at relatively lower and medium SNR regimes. Our observations show that the potential for error coefficient improvement increases with an increase in code rate. The reason is that low-rate codes usually have a small error coefficient due to a limited number of cosets $\Ci,i\in\B$ in these codes. Hence, a significant reduction in the error coefficient is hard to achieve. 

%Meanwhile, the benefit of following one decoding path, SD requires less memory and resources compared to SCLD(8).
%Discussion on the performance comparison based on the table below: 

%%%%%%%%%%%%%%%%%%%%%%%%%%%%%%%%%%%%%%%%%%%%%%%%%%%%%%%%%%%%%%%%%%%%%
\section{CONCLUSION} %AND FUTURE DIRECTIONS}

In this paper, we analyzed the formation of MWCs in CRC-polar codes and proposed two novel schemes: profile-shifted PAC codes and continuous CRC-polar codes. By reserving large-index coordinates in the rate profile, PS-PAC codes overcome PAC limitations, significantly reducing minimum-weight codewords and improving error correction across various rates and lengths. Inspired by PAC codes, CCRC-polar codes continuously utilize the intermediate remainder of CRC coding, enhancing performance for long codes.

\addtolength{\textheight}{-12cm}   % This command serves to balance the column lengths
                                  % on the last page of the document manually. It shortens
                                  % the textheight of the last page by a suitable amount.
                                  % This command does not take effect until the next page
                                  % so it should come on the page before the last. Make
                                  % sure that you do not shorten the textheight too much.

%%%%%%%%%%%%%%%%%%%%%%%%%%%%%%%%%%%%%%%%%%%%%%%%%%%%%%%%%%%%%%%%%%%%%%%%%%%%%%%%

%%%%%%%%%%%%%%%%%%%%%%%%%%%%%%%%%%%%%%%%%%%%%%%%%%%%%%%%%%%%%%%%%%%%%%%%%%%%%%%%

%%%%%%%%%%%%%%%%%%%%%%%%%%%%%%%%%%%%%%%%%%%%%%%%%%%%%%%%%%%%%%%%%%%%%%%%%%%%%%%%
%\Section*{APPENDIX}

\bibliographystyle{IEEEtran}
\bibliography{references} 
\end{document}